\documentclass{proc}

\begin{document}

\title{Data Mining Using High Performance Data Clouds: \protect\\ 
Experimental Studies Using Sector and Sphere}
\numberofauthors{2}
\author{
\alignauthor
Robert L Grossman \\
\affaddr{University of Illinois at Chicago} \\
\affaddr{and Open Data Group} \\
\alignauthor
Yunhong Gu \\
\affaddr{University of Illinois at Chicago} \\
}

\maketitle

\begin{abstract}
  We describe the design and implementation of a high performance
  cloud that we have used to archive, analyze and mine large
  distributed data sets.  By a cloud, we mean an infrastructure that
  provides resources and/or services over the Internet.  A storage
  cloud provides storage services, while a compute cloud provides
  compute services.  We describe the design of the Sector storage
  cloud and how it provides the storage services required by the
  Sphere compute cloud.  We also describe the programming paradigm
  supported by the Sphere compute cloud.  Sector and Sphere are
  designed for analyzing large data sets using computer clusters
  connected with wide area high performance networks (for example, 10+
  Gb/s).  We describe a distributed data mining application that we have
  developed using Sector and Sphere.  Finally, we describe some
  experimental studies comparing Sector/Sphere to Hadoop.
\end{abstract}

\vspace{1mm}
\noindent {\bf Categories and Subject Descriptors:} H.2.8 {[Database
  Management]}: Data mining, C.2.4 {[Computer-Communications Networks]}:
Distributed applications, D.4.3 {[Operating Systems]}: Distributed
file systems, D.4.1 {[Process Management]}:
Multiprocessing/multiprogramming/multitasking

\vspace{1mm}
\noindent
{\bf General Terms:} design, experimentation, measurement, performance

\vspace{1mm}

\section{Introduction}

Historically, high performance data mining systems have been designed
to take advantage of powerful, but shared pools of processors.
Generally, data is scattered to the processors, the computation is
performed using a message passing or grid services library, the
results are gathered, and the process is repeated by moving new data
to the processors.

This paper describes a distributed high performance data mining system
that we have developed called Sector/Sphere that is based on an
entirely different paradigm.  Sector is designed to provide long term
persistent storage to large datasets that are managed as distributed
indexed files.  Different segments of the file are scattered throughout
the distributed storage managed by Sector. Sector generally replicates
the data to ensure its longevity, to decrease the latency when
retrieving it, and to provide opportunities for parallelism.  Sector
is designed to take advantage of wide area high performance networks
when available.  

Sphere is designed to execute user defined functions in parallel using
a stream processing pattern for data managed by Sector.  We mean by
this that the same user defined function is applied to every data
record in a data set managed by Sector.  This is done to each segment
of the data set independently (assuming that sufficient processors are
available), providing a natural parallelism. The design of
Sector/Sphere results in data frequently being processed in place
without moving it.

To summarize, Sector manages data using distributed, indexed files; 
Sphere processes data with user-defined functions that operate
in a uniform manner on streams of data managed by Sector; Sector/Sphere
scale to wide area high performance networks using specialized network
protocols designed for this purpose.

In this paper, we describe the design of Sector/Sphere.  We also describe
a data mining application developed using Sector/Sphere that searches for
emergent behavior in distributed network data.  We also describe various
experimental studies that we have done using Sector/Sphere.  Finally, we 
describe several experimental studies comparing Sector/Sphere to Hadoop
using the Terasort Benchmark \cite{Borthakur:2007}, as well as a companion
benchmark we have developed called Terasplit that computes a split for
a regression tree.

This paper is organized as follows: Section 2 describes background and
related work.  Section 3 describes the design of Sphere.  Section 4
describes the design of Sector.  Section 5 describes the design of the
networking and routing layer.  Section 6 contains some experimental
studies.  Section 7 describes a Sector/Sphere application that we have
developed.  Section 8 is the summary and conclusion.

This paper is based in part on \cite{Grossman:FGCS08}.  In particular,
some of the introductory and background material is the same.  This
paper describes a later version of the Sector/Sphere system, describes
the Sector/Sphere system in more detail, and contains different
experimental results.

\section{Background and Related Work}
\label{background}

By a cloud, we mean an infrastructure that provides resources and/or
services over the Internet.  A {\em storage cloud} provides storage
services (block or file based services); a {\em data cloud} provides
data management services (record-based, column-based or object-based
services); and a {\em compute cloud} provides computational services.
Often these are layered (compute services over data services over
storage service) to create a stack of cloud services that serves as a
computing platform for developing cloud-based applications.

Examples include Google's Google File System (GFS), BigTable and
MapReduce infrastructure \cite{Dean:04}, \cite{Ghemawat:03}; Amazon's
S3 storage cloud, SimpleDB data cloud, and EC2 compute cloud
\cite{Amazon:WS07}; and the open source Hadoop system
\cite{Borthakur:2007}, \cite{Hbase:2007}.

In this section, we describe some related work in high performance and
distributed data mining.  For a  recent survey of high performance and
distributed data mining systems, see \cite{NGDM:2007}.

By and large, data mining systems that have been developed to date for
clusters, distributed clusters and grids have assumed that the
processors are the scarce resource, and hence shared.  When processors
become available, the data is moved to the processors, the computation
is started, and results are computed and returned \cite{Foster:Grid2}.
In practice with this approach, for many computations, a good portion
of the time is spent transporting the data.  

In contrast, the approach taken here by Sector/Sphere is to store the
data persistently and to process the data in place when possible.  In
this model, the data waits for the task or query.  The storage clouds
provided by Amazon's S3 \cite{Amazon:S3}, the Google File System
\cite{Ghemawat:03}, and the open source Hadoop Distributed File System
(HDFS) \cite{Borthakur:2007} support this model.

MapReduce and Hadoop and their underlying file systems GFS and HDFS
are specifically designed for racks of computers in data centers.
Both systems use information about clusters and racks to position
file blocks and file replicas.  This approach does not work well with
loosely coupled distributed environments, such as those that Sector
targets.

To date, work on storage clouds \cite{Ghemawat:03, Borthakur:2007,
Amazon:S3} has assumed relatively small bandwidth between the
distributed clusters containing the data.  In contrast, the Sector
storage cloud described in Section~\ref{section:sector} is designed
for wide area, high performance 10 Gb/s networks and employs
specialized protocols, such as UDT \cite{Grossman:CN2007}, to utilize
the available bandwidth on these networks.  

Sector is also designed for loosely coupled distributed systems that are
managed with a peer-to-peer architecture, while storage
clouds such as GFS and HDFS are designed for more tightly coupled
systems that are managed with a centralized master node.

In addition, Sector assumes that the data is divided into files, while
GFS and HDFS divide the data into blocks that are scattered across
processors.  For example, as usually configured Sector processes a 1
TB file using 64 chunks, each of which is a file, while HDFS process
the same data using 8,192 chunks, each of which is a block.  (The
default block size for HDFS is 64 MB.  We increased this to 128 MB for
the experiments described below, which improved the Hadoop
experimental results.)

The most common way to code data mining algorithms on clusters and
grids is to use message passing, such as provided by the MPI library
\cite{Gropp:1999}, or to use grid libraries and services, such as
globus-url-copy to scatter and gather data and programs and
globus-job-run to run programs \cite{Foster:Grid2}.

The most common way to compute over GFS and HDFS storage clouds is to
use MapReduce \cite{Dean:04}.  With MapReduce: i) relevant data is
extracted in parallel over multiple nodes using a common ``map''
operation; ii) the data is then transported to other nodes as required
(this is referred to as a shuffle); and, iii) the data is then
processed over multiple nodes using a common ``reduce'' operation to
produce a result set.  In contrast, the Sphere compute cloud described
in Section~\ref{section:sphere} allows arbitrary user defined
operations to replace both the map and reduce operations.  In
addition, Sphere uses the same specialized network transport protocols
\cite{Grossman:CN2007} that Sector uses so that any transfer of data
required by Sphere's user defined functions can be transferred
efficiently over wide area high performance networks.

\begin{figure}
\begin{center}
\begin{tabular}{|c|c|c|} \hline
Application 1 & $\cdots$ & Application $n$ \\ \hline
\multicolumn{3}{|c|}{Cloud-based Compute Services} \\ \hline
\multicolumn{3}{|c|}{Cloud-based Data Services} \\ \hline
\multicolumn{3}{|c|}{Cloud-based Storage Services} \\ \hline
\end{tabular}
\end{center}
\caption{A data stack for a cloud consists of three layered
services as indicated.}
\label{figure:datastack}
\end{figure}

\section{Design of Sphere}
\label{section:sphere}

\subsection{Overview}

The Sphere Compute Cloud is designed to be used with the Sector 
Storage Cloud.  Sphere is designed so that certain specialized, but
commonly occurring, distributed computing operations can be done
very simply.  Specifically, if a user defines a function $p$ on
a distributed data set $a$ managed by Sector, then invoking the
command
\begin{verbatim}
  sphere.run(a, p);
\end{verbatim}
applies the user defined function $p$ to each data record in the
dataset $a$.  In other words, if the dataset $a$ contains 
$100,000,000$ records $a[i]$, then the Sphere command above
replaces all the code required to read and write the array $a[i]$
from disk, as well as the loop:
\begin{verbatim}
  for (int i = 0, i < 100000000; ++i)
    p(a[i]);
\end{verbatim}

The Sphere programming model is a simple example of what is commonly
called a stream programming model.  Although this model has been used
for some time, it has recently received renewed attention due to its
use by the general purpose GPU (Graphics Processing Units) community
(GPGPU community) \cite{Owens:2005} and by the popularization of the
MapReduce \cite{Dean:04} special case, which restricts attention to
data of the form $[\mbox{key},\ \mbox{value}]$ and to two user defined
functions (Map and Reduce).

Large data sets processed by Sphere are assumed to be broken up into
several files.  For example, the Sloan Digital Sky Survey dataset
\cite{Szalay:Science01} is divided up into 64 separate files, each 
about 15.6 GB in size.  The files are named sdss1.dat, $\ldots$,
sdss64.dat.  

Assume that the user has a written a function called
{\em findBrownDwarf} that given a record
in the SDSS dataset, extracts candidate Brown Dwarfs.
Then to find brown dwarfs in the Sloan dataset, one uses the following
Sphere code:
\begin{verbatim}
  Stream sdss;
  sdss.init(...);   //init with 64 sdss files 
  Process* myproc = Sector::createJob();
  myproc->run(sdss, "findBrownDwarf");
  myproc->read(result);
\end{verbatim}
With this code, Sphere uses Sector to access the required SDSS files,
uses an index to extract the relevant records, and for each record
invokes the user defined function {\em findBrownDwarf}.  Parallelism
is achieved in two ways.  First, the individual files can be processed
in parallel.  Second, Sector is typically configured to create replicas
of files for archival purposes.  These replicas can also be processed
in parallel.  

An important advantage provided by a system such as Sphere is that
often data can be processed in place, without moving it.  In contrast,
a grid system generally transfers the data to the processes prior
to processing \cite{Foster:Grid2}.

\begin{figure*}
\centering
\includegraphics[scale=0.9]{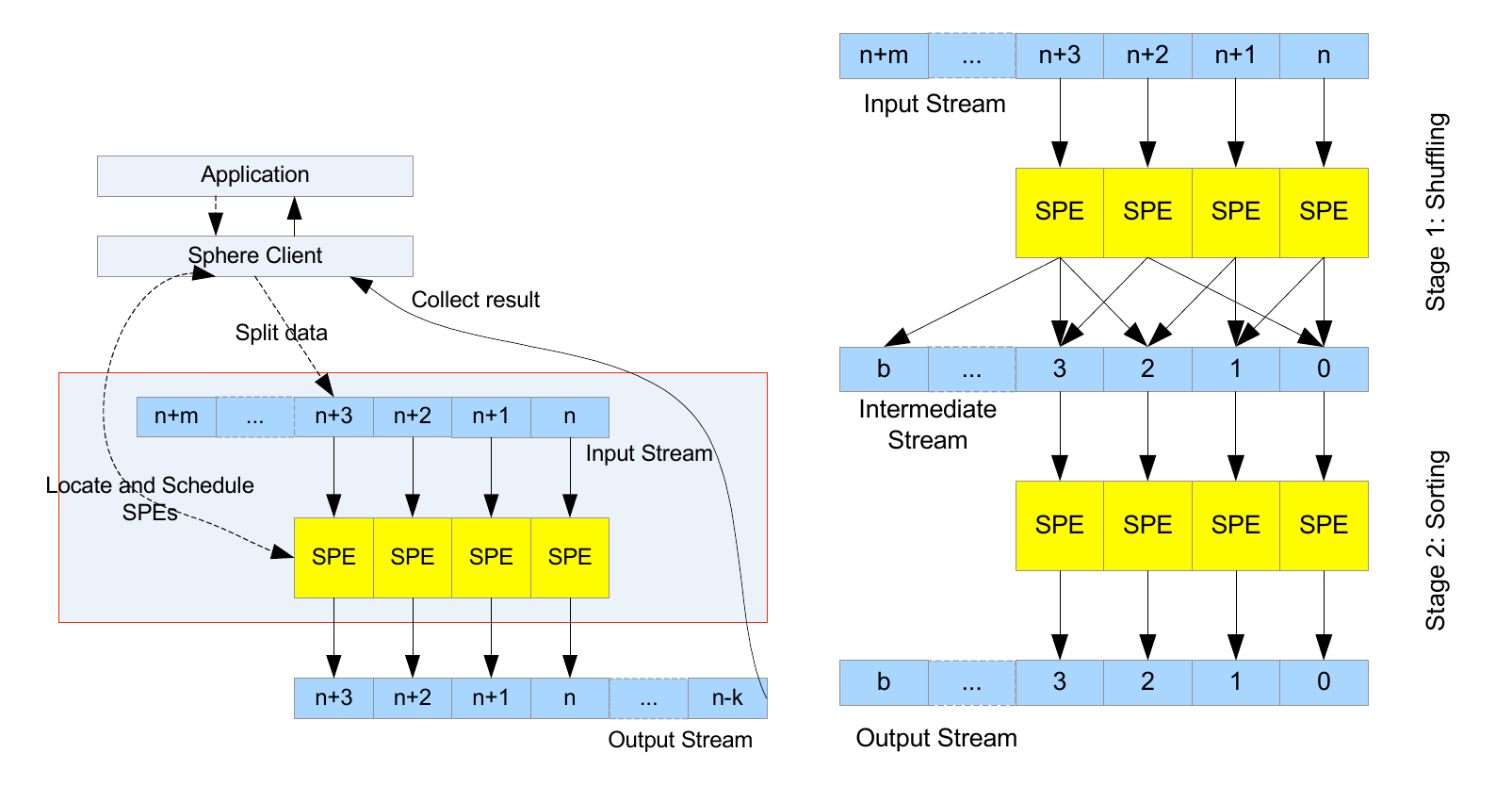}
\caption{This figure illustrates how Sphere operators process
Sphere streams over distributed Sphere Processing Elements (SPE).}
\label{figure:sphere-design}
\end{figure*}

\subsection{Sphere Computing Model}

The computing model used by Sphere is based upon the following
concepts.  A Sphere {\em dataset} consists of one or more physical
{\em files}.  Computation in Sphere is done by user defined functions
({\em Sphere operator} that take a Sphere {\em stream} as input and
produce a Sphere stream as output.  Sphere streams are split into one
or more {\em data segments} that are processed by Sphere servers, which
are called {\em Sphere Processing Elements} or SPE.  Sphere data
segments can be a data record, a collection of data records, or a
file.  See Figure~\ref{figure:sphere-design}.

When a Sphere function processes a stream, the resulting stream can be
returned to the Sector node where it originated, written to a local
node, or ``shuffled'' to a list of nodes, depending upon how the
output stream is defined.

The SPE is the major Sphere service and it is started by a Sphere
server in response to a request from a Sphere client. Each SPE is
based on a user-defined function (Sphere operator).  The Sphere
operator is implemented as a dynamic library and is stored on the
server's local disk, which is managed by the Sector server. For
security reasons, uploading such library files to a Sector server is
limited. A library file resides on a Sector server only if the Sphere
client program has write access to the particular Sector server or the
server's owner has voluntarily downloaded the file. Sector's replica
service is disabled for Sphere operators.

Once the Sphere server accepts the client's request, it starts an SPE
and binds it to the local Sphere operator. The SPE runs in
a loop and consists of the following four steps:

\begin{enumerate}

\item The SPE accepts a new data segment from the client, which
  contains the file name, offset, number of rows to be processed, and
  additional parameters.

\item The SPE reads the data segment and its record index from local
  disk or from a remote disk managed by Sector.

\item For each data segment (single data record, group of data records, or
  entire data file), the Sphere operator processes the
  data segment and writes the result to a temporary buffer. In addition,
  the SPE periodically sends acknowledgments to the client about the
  progress of the processing.

\item When the data segment is completely processed, the SPE sends an
  acknowledgment to the client and writes the results to the
  appropriate destinations, as specified in the output stream.  If
  there are no more data segments to be processed, the client closes
  the connection to the SPE, and the SPE is released.

\end{enumerate}

Sphere assigns SPEs to streams as follows:

\begin{enumerate}

\item The stream is first divided into data segments.  This is done
  roughly as follows.  The total data size $S$ and the total number of
  records $R$ is computed.  Say the number of SPEs available for the
  job is $N$.  Roughly speaking, the number of records that equals
  $S/N$ should be assigned to each SPE.  The user specifies a minimum
  and maximum data size $S_{min}$ and $S_{max}$ that should be
  assigned to each processor.  If $S/N$ is between these user defined
  limits, the associated number of records is assigned to each
  SPE. Otherwise the nearest boundary $S_{min}$ or $S_{max}$ is used
  instead to compute the required number of records to assign to each
  SPE.

\item Once the stream is segmented into data segments of the
  appropriate size, each data segment is assigned to a SPE on the same
  machine whenever possible.

\item Data segments from the same file are not processed at the same
  time, unless not doing so would result in an idle SPE.

\end{enumerate}

\section{Design of Sector}
\label{section:sector}

\begin{figure}
\begin{center}
\includegraphics[scale=0.4]{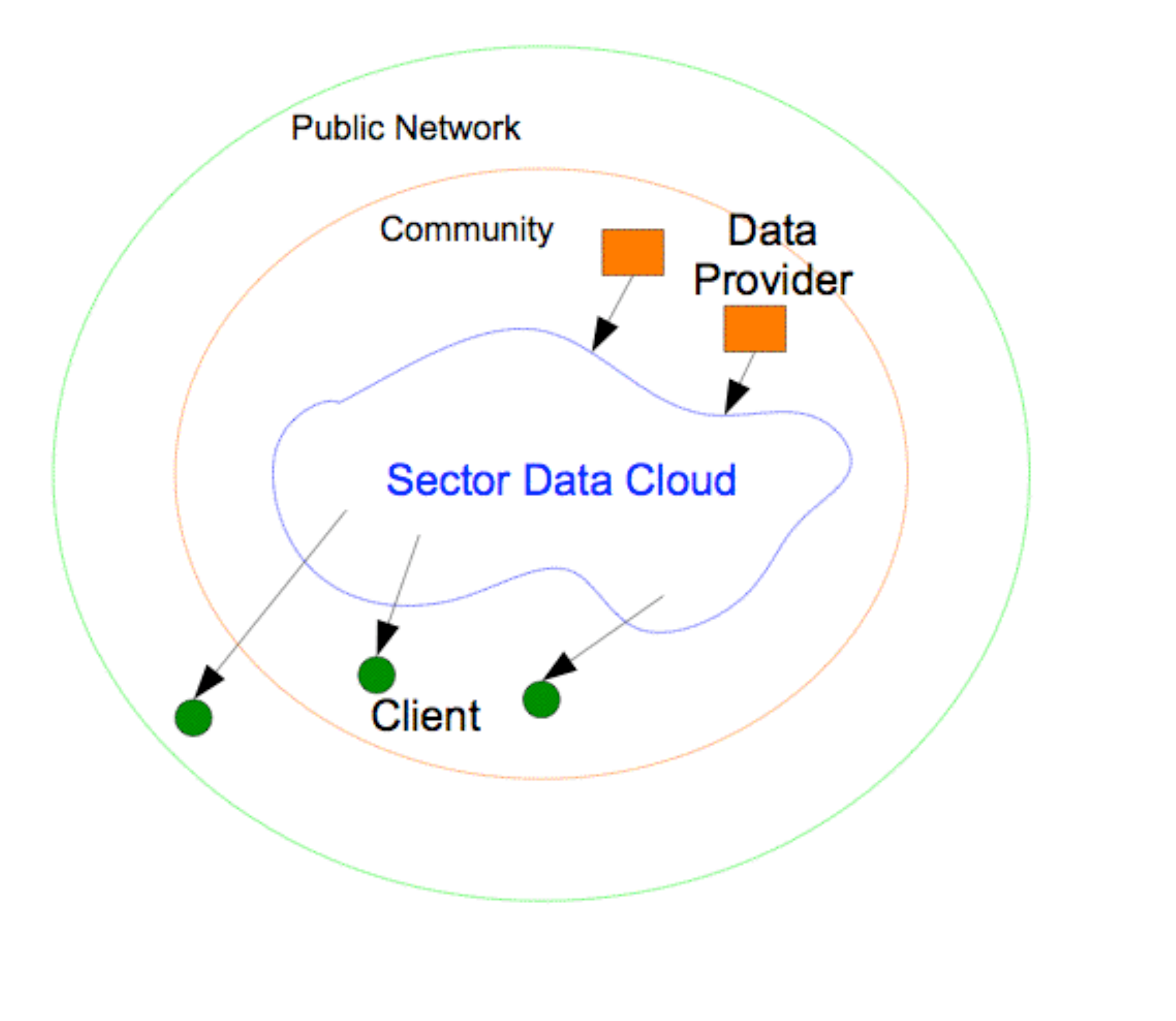}
\end{center}
\caption{With Sector, only users in a community who have been added to
  the Sector access control  can write data into Sector.  On the
  other hand, any member of the community or of the public can read
  data, unless additional restrictions are imposed.}
\label{figure:sector-access}
\end{figure}

Sector is the underlying storage cloud that provides persistent
storage for the data required by Sphere and manages the data for
Sphere operations. Since some portions of Sector have been described
previously \cite{Grossman:eScience06}, we present just a brief
summary here. Sector is not a file system per se, but rather provides
services that rely in part on the local native file systems.

The core requirements for Sector are:

\begin{enumerate}

\item Sector provides long term archival storage and access for large
distributed datasets.

\item Sector is designed to utilize the bandwidth available
on wide area high performance networks.

\item Sector supports a variety of different routing and network protocols.

\item Sector is designed to support a community of users, not all of
  whom may have write access to the Sector infrastructure.

\end{enumerate}

Sector uses replication in order to safely archive data. It monitors
the number of replicas, and, when necessary, creates additional
replicas at a random location. The number of replicas of each file is
checked once per day. The choice of random location leads to
uniform distribution of data over the whole system.

Sector takes advantage of wide area, high performance networks by
using specialized network transport protocols such as UDT
\cite{Grossman:CN2007}. Sector also caches data connections. Therefore,
frequent data transfers between the same pair of nodes do not need
to set up a data connection every time. This reduces the connection
setup overhead.

Sector has separate layers for routing and transport and interfaces
with these layers through well defined APIs. In this way, it is
relatively straightforward to use other routing or network
protocols. In addition, UDT is designed in such a way that a variety
of different network protocols can be used simply by linking in one
of several different libraries \cite{Grossman:CN2007}.

Sector's security mechanism is enabled by Access Control List (ACL).
While data read is open to the general public, write access to the
Sector system is controlled by ACL, as the client's IP address must
appear in the server's ACL in order to upload data to that
particular server.  See Figure~\ref{figure:sector-access}.

Sector was designed to provide persistent storage services for data
intensive applications that involve mining multi-terabyte datasets
accessed over wide area 10 Gb/s networks.

As an example, Sector is used to archive and to distribute the Sloan
Digital Sky Survey (SDSS) to astronomers around the world.  Using
Sector, the SDSS BESTDR5 catalog, which is about 1.3TB when
compressed, can be transported at approximately 8.1 Gb/s over a 10
Gb/s wide area network with only 6 commodity servers
\cite{Grossman:BWC2006}.

Sector assumes that large datasets are divided into multiple files,
say file01.dat, file02.dat, etc.  It also assumes that each file is
organized into records.  In order to randomly access a record in
the data set, each data file in Sector has a companion index file,
with a post-fix of ``.idx''.  Continuing the example above, there
would be index files file01.dat.idx, file02.dat.idx, etc.  The data
file and index file are always co-located on the same node.  Whenever
Sector replicates the data file, the index file is also replicated.

The index contains the start and end positions (i.e., the offset and
size) of each record in the data file.  For those data files without
an index, Sphere can only process them at the file level, and the user must
write a function that parses the file and extracts the data.

A Sector client accesses data using Sector as follows:

\begin{enumerate}

\item The Sector client connects to a known Sector server S, and
  requests the locations of an entity managed by Sector using the
  entity's name.

\item The Sector Server S runs a look-up inside the server network
  using the services from the routing layer and returns one or more
  locations to the client.  In general, an entity managed by Sector is
  replicated several times within the Sector network.  The routing
  layer can use information involving network bandwidth and latency to
  determine which replica location should be provided to the client.

\item The client requests a data connection to one or more servers on
  the returned locations using a specialized Sector library designed
  to provide efficient message passing between geographically
  distributed nodes.  The Sector library used for messaging uses a
  specialized protocol developed for Sector called the Group Messaging
  Protocol.

\item All further requests and responses are performed using a
  specialized library for high performance network transport called
  UDT \cite{Grossman:CN2007}.  UDT is used over the data connection
  established by the message passing library.

\end{enumerate}

\begin{figure}
\begin{center}
\begin{tabular}{|c|c|c|} \hline
Sector Application 1 & $\cdots$ & Sector Application $n$ \\ \hline
\multicolumn{3}{|c|}{File Location and Access Services} \\ \hline
\multicolumn{3}{|c|}{Distributed Storage Services} \\ \hline
\multicolumn{3}{|c|}{Routing Services} \\ \hline
\multicolumn{3}{|c|}{Network Transport Services} \\ \hline
\end{tabular}
\end{center}
\caption{Sector consists of several layered services.}
\label{figure:sectorstack}
\end{figure}

\section{Design of Networking Layer}
\label{section:networking}

As mentioned above, Sector is designed to support a variety of
different routing and networking protocols.  The version used for the
experiments described below are designed to support large distributed
datasets, with loose management provided by geographically distributed
clusters connected by a high performance wide area network.  With this
configuration, a peer-to-peer routing protocol (the Chord protocol
described in \cite{Stoica:2001}) is used so that nodes can be easily
added and removed from the system.

The next version of Sector will support specialized routing protocols
designed for wide area clouds with uniform bandwidth and approximately
equal RTT between clusters, as well as non-uniform clouds in which
bandwidth and RTT may vary widely between different clusters of the
cloud.

Data transport within Sector is done using specialized network
protocols.  In particular, data channels within Sector use high
performance network transport protocols, such as UDT
\cite{Grossman:CN2007}.  UDT is a rate-based application layer network
transport protocol that supports large data flows over wide area high
performance networks.  UDT is {\em fair} to several large data flows
in the sense that it shares bandwidth equally between them.  UDT is
also {\em friendly} to TCP flows in the sense that it backs off when
congestion occurs, enabling any TCP flows sharing the network to use
the bandwidth they require.

Message passing with Sector is done using a specialized network transport
protocol that we developed for this purpose called the Group Messaging
Protocol or GMP.

\section{Experimental Studies}

\subsection{Experimental Setup}

The wide area experiments use 6 servers: two are in Chicago, Illinois;
two are in Greenbelt, Maryland; and two are in Pasadena, California.
The wide area servers have double dual-core 2.4 GHz Opteron processors,
4GB RAM, 10GE MyriNet NIC, and 2TB of disk.

The round trip time (RTT) between the servers in Greenbelt and Chicago
is 16ms.  The RTT between Chicago and Pasadena is 55ms. The servers in
Greenbelt and Pasadena are networked through Chicago and therefore the
RTT is 71 ms.  All the servers are connected with 10 Gb/s networks.

The local area experiments use 8 servers that have dual
4-core 2.4 GHz Xeon processors, 16GB RAM, 10GE MyriNet NIC, and 5.5TB of
disk.  Note that the servers for the local area experiments are newer
than those used for the wide area experiments.

The version of Hadoop used for the experimental studies was version
0.16.0.  The Java(tm) version was 1.6.0, the Java(tm) SE Runtime
Environment was build 1.6.0-b105; the Java HotSpot(tm) 64-Bit Server
VM was build 1.6.0-b105, mixed mode.

\subsection{Hadoop vs Sphere - Geographically Distributed Locations}

In this section, we perform the tests using Terasort but this time
using six servers that are geographically distributed. Two of the
servers are in Chicago Illinois, two are in Pasadena, California, and
two are in Greenbelt, Maryland.  All the servers are connected with a
10 Gb/s network. 

Table~\ref{fig:exp-wide-area} compares the performance of the Terasort
benchmark (sorting 10GB data on each node, 100-byte record with
10-byte key) using both Hadoop and Sphere.

To understand the performance of Sector/Sphere for typical data mining
computations, we developed a benchmark that we call {\em Terasplit}.
Terasplit takes data that has been sorted, for example by Terasort,
and computes a single split for a tree based upon entropy
\cite{Breiman:1984}.  Although Terasplit benchmarks could be developed
for multiple clients, the version we use for the experiments here read
(possibly distributed) data into a single client to compute the split.
Table~\ref{fig:exp-wide-area} also compares the performance of
Sector/Sphere and Hadoop for the Terasplit benchmark.

\begin{table*}
\begin{center}
\begin{tabular}{|p{1.50in}|p{0.50in}|p{0.50in}|p{0.50in}|p{0.50in}|
p{0.50in}|p{0.50in}|}\hline
{\bf Nodes Used} & {\bf 1} & {\bf 1-2}  & {\bf 1-3} & {\bf 1-4} &  
{\bf 1-5} & {\bf 1-6} \\ \hline 
Size of Dataset (GB) & 10 & 20 & 30 & 40 & 50 & 60 \\ \hline
Locations & \multicolumn{2}{|c|}{1} & \multicolumn{2}{|c|}{2} & 
\multicolumn{2}{|c|}{3} \\ \hline \hline
Hadoop Terasort  & 2312 & 2401 & 2623 & 3228 & 3358 & 3532 \\ \hline
Sphere Terasort  & 905 & 980 & 1106 & 1260  & 1401 & 1450  \\ \hline \hline
Hadoop Terasplit & 460 & 623 & 860 & 1038 & 1272 & 1501 \\ \hline
Sphere Terasplit  & 110 & 320 & 422 & 571  & 701 & 923  \\ \hline  \hline
Total Hadoop  & 2772 & 3024 & 3483 & 4266 & 4657 & 5033 \\ \hline
Total Sphere  & 1015 & 1300 & 1528 & 1831  & 2102 & 2373  \\ \hline  \hline
Speedup Terasort  & 2.6 & 2.5 & 2.4 & 2.6 & 2.4 & 2.4 \\ \hline
Speedup Terasplit  & 4.2 & 1.9 & 2.0 & 1.8  & 1.8 & 1.6  \\ \hline
Speedup total & 2.7 & 2.3 & 2.3 & 2.3 & 2.2 & 2.1 \\ \hline 
\end{tabular}
\end{center}
\caption{This table compares the performance of Sphere and Hadoop
  sorting a 10GB file on each of six nodes that are distributed
  over a wide area network using the Terasort benchmark.  The performance
  using the Terasplit benchmark is also reported, as is the total for 
  Terasort plus Terasplit.  The speedup of Sphere compared to Hadoop
  is reported for the Terasort and Terasplit benchmarks, as well as the total
  of the two.  Nodes 1 and 2 are located in Chicago;
  nodes 3 and 4 are located in Pasadena; nodes 5 and 6 are 
  located in Greenbelt. All measurements are in seconds. The nodes were double dual-core 
  2.4 GHz Opteron processors with 4 GB of memory. N.B. Different types of
  servers  were used for the local and wide area tests.  }
\label{fig:exp-wide-area}
\end{table*}

\subsection{Hadoop vs Sphere - Single Location}

\begin{table*}
\begin{center}
\begin{tabular}{|p{1.50in}|p{0.50in}|p{0.50in}|p{0.50in}|p{0.50in}|
p{0.50in}|p{0.50in}|p{0.50in}|p{0.50in}|}\hline {\bf Nodes Used} &
{\bf 1} & {\bf 1-2}  & {\bf 1-3} & {\bf 1-4} & {\bf 1-5} & {\bf 1-6}
& {\bf 1-7} & {\bf 1-8} \\ \hline Size of Dataset (GB) & 10 & 20 &
30 & 40 & 50 & 60 & 70 & 80 \\ \hline Hadoop Terasort  & 645 & 766 &
768 & 773 & 815 & 882 & 901 & 1000 \\ \hline Sphere Terasort  & 408
& 409 & 410 & 429 & 430 & 436 & 440 & 443 \\ \hline \hline Hadoop
Terasplit & 141 & 266 & 410 & 544 & 671 & 901 & 1133 & 1250 \\ \hline 
Sphere Terasplit  & 96 & 221 & 350 & 462 & 560 & 663 & 754 & 855 \\ \hline\hline 
Total Hadoop & 786 & 1032 & 1178 & 1317 & 1486 & 1784 & 2034 & 2250 \\ \hline 
Total Sphere  & 504 & 630 & 760 & 891 & 990 & 1099 & 1194 & 1298 \\ \hline
\hline Speedup Terasort  & 1.6 & 1.9 & 1.9 & 1.8 & 1.9 & 2.0 & 2.0 & 2.3\\ \hline 
Speedup Terasplit  & 1.5 & 1.2 & 1.2 & 1.2 & 1.2 & 1.4 & 1.5 & 1.5 \\ \hline 
Speedup total & 1.6 & 1.6 & 1.6 & 1.5 & 1.5 & 1.6 & 1.7 & 1.7 \\ \hline
\end{tabular}
\end{center}
\caption{This table compares the performance of Sphere and Hadoop
  sorting a 10GB file on each of eight nodes, all of which are located
  on a single rack. All measurements are in seconds. The nodes were dual quad core 
  2.4 GHz Xeon processors with 16 GB of memory.}
\label{fig:exp-local}
\end{table*}

In this section we describe some comparisons between Sphere and Hadoop
\cite{Borthakur:2007} on 8-node Linux cluster in a single location.
As for the wide area experiments, we ran both the Terasort and Terasplit
benchmarks.

The file generation required 212 seconds per file per node for Hadoop,
which is a throughput of 440Mb/s per node. For Sphere, the file
generation required 68 seconds per node, which is a throughput of
1.1Gb/s per node.

Both Hadoop and Sphere scale very well with respect to the Terasort
and Terasplit benchmarks, as the table indicates.
Sphere is about 1.6--2.3 times faster than Hadoop as measured by the
Terasort benchmark and about 1.2-1.5 times faster as measured
by the Terasplit benchmark.  

Although we expected Sector/Sphere to be faster for the wide area
experiments, we did not expect to see such a difference for
the local area experiments.  This may be due in part to our ability
to tune Sphere more proficiently than we can tune Hadoop.  Also,
we noted that Hadoop performed better on clusters employing 1 Gb/s
NICs than 10 Gb/s NICs.  Sector/Sphere has been tested extensively
using 10 Gb/s NICs and Hadoop may not have been.

\subsection{Discussion}

As mentioned above, for the Terasort benchmark, Sector/Sphere only uses one of
the 4 available cores, while Hadoop uses all 4 cores.  For this reason,
the Terasort performance is not exactly comparable.  

Note that Sector/Sphere provides a performance improvement of approximately
2.4--2.6 over a wide area network compared to Hadoop as measured by the
Terasort benchmark, a performance improvement of
1.6--1.8 as measured by the Terasplit benchmark, and a performance
improvement of 2.1-2.3 for the Terasort$+$Terasplit benchmark.

For local area clusters, Sector/Sphere is about 
1.6--2.3 times faster as measured by the Terasort benchmark
and 1.2--1.5 times faster as measured by the Terasplit benchmark.
As mentioned above, this difference may be due to the fact that
Hadoop has not been tuned to work with 10 Gb/s NICs.

Note that from the experimental studies reported in
Table~\ref{fig:exp-wide-area}, both Sector/Sphere scale to wide area
networks.  Specifically, note that Sector/Sphere scales to four nodes
in two distributed locations over a network with a RTT of 16 ms with a
performance impact of approximately 41\%,  (for Hadoop, the impact is
also approximately 41\%).  For three locations, with RTT of 16 ms, 55
ms and 71 ms between, the performance impact is approximately 82\%,
while for Hadoop the impact is  approximately 67\%.

\subsection{Availability and Repeatability}

Version 1.4 of Sector Sphere was used for the experimental studies
described here.  This version of Sector (as well as previous versions)
is available from the Source Forge web site \cite{Sector:v1.4}.

The Terasort benchmark is available from \cite{Borthakur:2007}.  The
Terasplit benchmark will be available with the next release of Sector
\cite{Sector:v1.4}; in the interim, it can be downloaded from
\cite{LargeDataArchive:2008}.

The Angle data set (used in the application below) is available from
the Large Data Archive \cite{LargeDataArchive:2008}.

With the Sector/Sphere software from Source Forge, the Terasort and
Terasplit benchmarks, and the Angle datasets from the Large Data
Archive, the experiments may all be repeated.  The results may vary
somewhat depending upon the specific servers used, the networks
connecting them, and the other network traffic present.

\section{Sphere Applications}

We have built several applications with Sector and Sphere.  In this
section, we describe one of them.

\subsection{Angle}

Angle is a Sphere application that identifies anomalous or suspicious
behavior in TCP packet data that is collected from multiple,
geographically distributed sites.   Angle
contains Sensor Nodes that are attached to the commodity Internet and
collect IP data. Connected to each Sensor Node on the commodity
network is a Sector node on a wide area high performance network.  The
Sensor Nodes zero out the content, hash the source and destination IP
to preserve privacy, package moving windows of anonymized packets in
pcap files \cite{Beale:2007} for further processing, and transfer
these files to its associated Sector node. Sector services are used to
manage the data collected by Angle and Sphere services are used to
identify anomalous or suspicious behavior.

Angle Sensors are currently installed at four locations: the
University of Illinois at Chicago, the University of Chicago, Argonne
National Laboratory and the ISI/University of Southern California.
Each day, Angle processes approximately 575 pcap files totaling
approximately 7.6GB and 97 million packets.  To date, we have
collected approximately 300,000 pcap files.

Briefly, Angle Sensor nodes collects IP data, anonymizes the IP data,
and produces pcap files that are then managed by Sector.  Sphere
aggregates the pcap files by source IP (or other specified entity) and
computes files containing features.  

Sphere is also used for processing
the feature files to identify emergent behavior.  This is done in several
ways.  One way is for Sphere to aggregate feature files into temporal windows,
$w_1$, $w_2$, $w_3$, $\ldots$, where each window is length $d$.   For each
window $w_j$, clusters are computed with centers $a_{j,1}$, $a_{j,2}$,
$a_{j,k}$ and the temporal evolution of these clusters is used to identify
certain clusters called emergent clusters.  For example, if the clusters
are relatively stable for windows $w_1$, $w_2$, $\ldots$, $w_{\alpha}$, but
there is statistically significant change in the clusters in $w_{\alpha+1}$,
then one or more clusters from window $w_{\alpha+1}$ can be identified.
These clusters are called {\em emergent clusters}.  

\begin{figure}[h]
\centering
\includegraphics[scale=0.4]{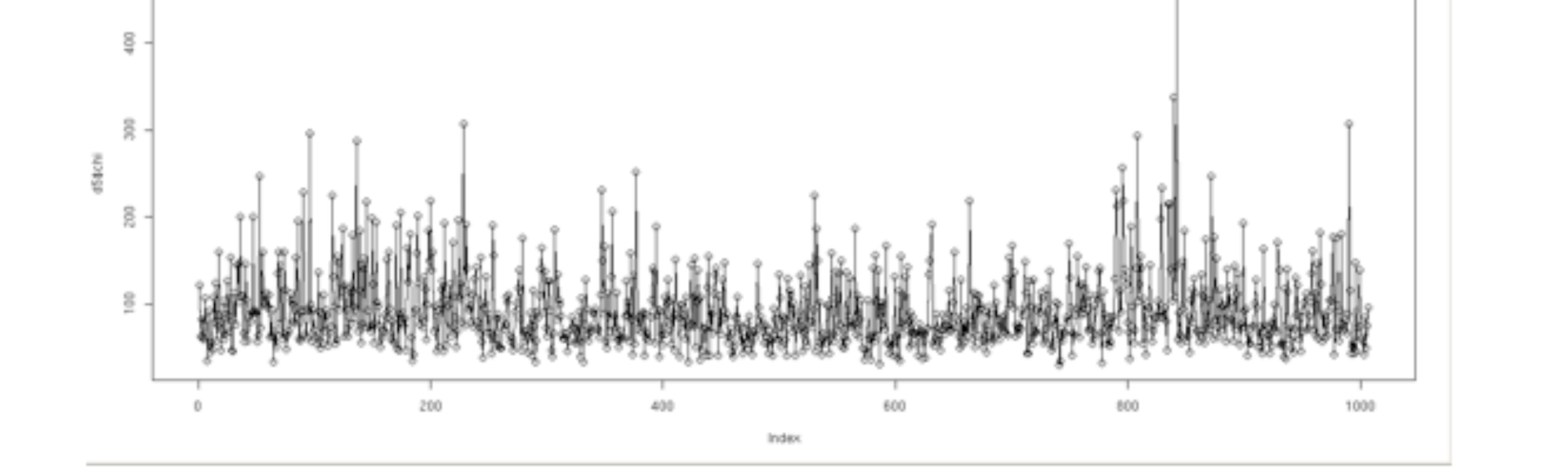}
\caption{The graph above shows how the cluster centers move from one
  ten minute window to another as measured by the statistic
  $\delta_j$.}
\label{fig:centers-10min}
\end{figure}

\begin{figure}[h]
\centering
\includegraphics[scale=0.4]{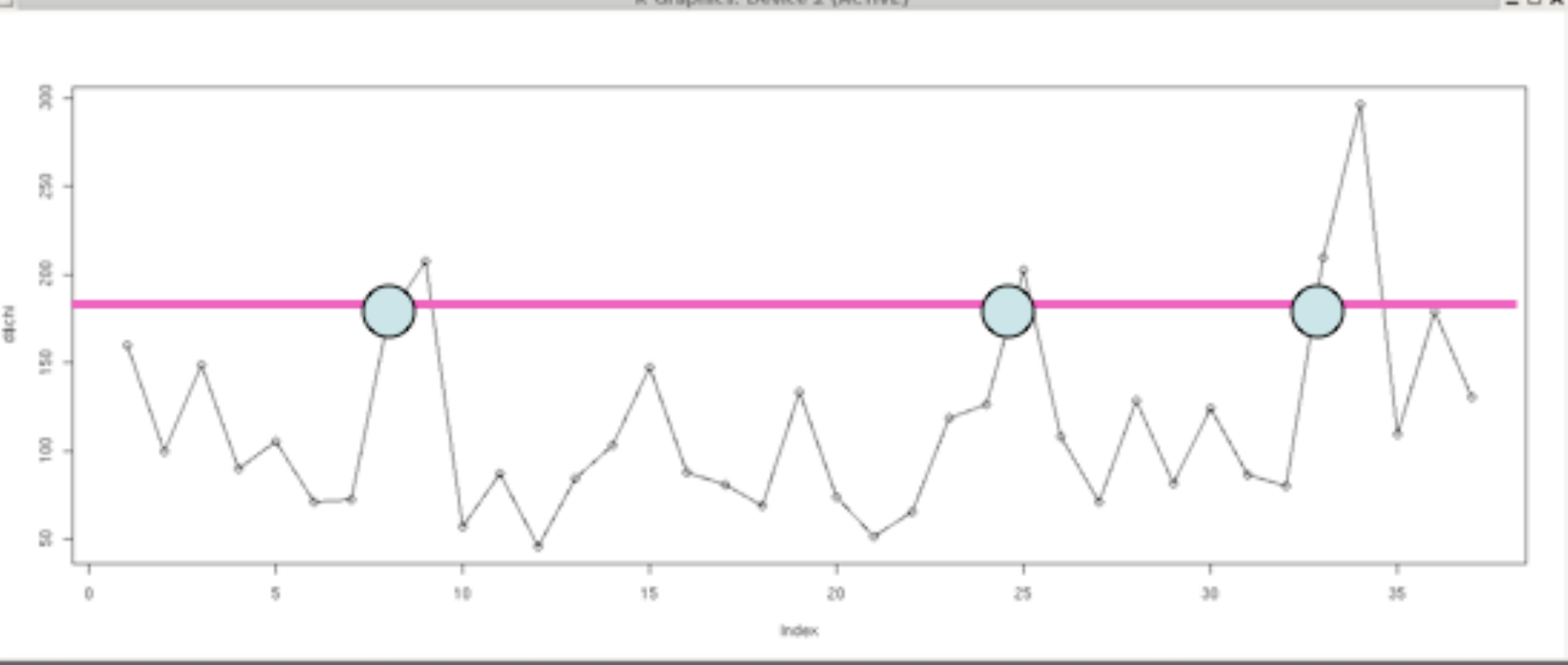}
\caption{The graph above shows how the cluster centers move from one
  1-day window to another as measured by the statistic
  $\delta_j$.  Emergent clusters were identified for the three days
  indicated and used as a basis for scoring functions.}
\label{fig:centers-1day}
\end{figure}

The following simple statistic can be used
$$ \delta_j = \sum_{i=1}^k \left(\min_{n\neq m} ||a_{j, n}- a_{j+1, m}||^2\right). $$
Figure~\ref{fig:centers-10min} shows this graph for windows of length
$d$ equals $10$ minutes.  Notice that the statistic $\delta_j$ is
quite choppy.  On the other hand, Figure~\ref{fig:centers-1day}
shows the same statistic for windows of length $d$ equals $1$ day.

Given one or more emergent clusters, a simple scoring function can be
used to identify feature vectors with emergent behavior.  For example,
if $\lambda_k$ are constants that sum to $1$, $a_k$ is the center of
an emergent cluster and $\sigma_k^2$ is its variance, then the
following score can be used to score feature vectors $x$
$$ \rho(x) = \max_{k} \rho_k(x)$$
$$ \rho_k(x) = \theta_k \exp\left({{-\lambda_k^2 ||x-a_k||^2}
\over{2 \sigma_k^2}}   \right),$$
where the max is over emergent clusters $k$.

See \cite{Grossman:NGDM2007} for more details.

Table~\ref{fig:shere-exp} shows the performance of Sector and Sphere
when computing cluster models as described above from distributed pcap
files. In this table, the work load varies from 1 to 300,000 distributed
pcap files.  This corresponds to approximately 500 to 100,000,000
feature vectors (each pcap file results in one file of features, which
are then aggregated and clustered, but a feature file can contain
various numbers of different feature vectors).

\begin{table}
\begin{center}
\begin{tabular}{|p{1.275in}|p{0.80in}|p{0.75in}|} \hline 
{\bf Number records} & {\bf Number of Sector Files} & {\bf Time} \\ \hline
500 & 1 & 1.9 s \\ \hline
1000 & 3 & 4.2 s \\ \hline
1,000,000 & 2850 & 85 min \\ \hline
100,000,000 & 300,000 & 178 hours \\ \hline
\end{tabular}
\end{center}
\caption{The time spent clustering using Sphere scales as the number of
files managed by Sector increases.}
\label{fig:shere-exp}
\end{table}

\section{Summary and Conclusion}

In this paper, we have described a cloud-based infrastructure designed
for data mining large distributed data sets over clusters connected
with high performance wide area networks.  Sector/Sphere is open
source and available through Source Forge.  We have used it as a basis
for several distributed data mining applications.

The infrastructure consists of the Sector storage cloud and the Sphere
compute cloud.  We have described the design of Sector and Sphere and
showed through experimental studies that Sector/Sphere can process
large datasets that are distributed across the continental U.S.  with
a performance penalty of approximately 80\% compared to the time
required if all the data were located on a single rack.  Sector/Sphere
utilize a specialized networking layer to achieve this performance.

We have also described a Sector/Sphere application to detect emergent
behavior in network traffic and showed that for this application
Sector/Sphere can compute clusters on over 300,000 distributed files.

Finally, we performed experimental studies on a wide area testbed and
demonstrated that Sector/Sphere is approximately 2.4--2.6 times faster
than Hadoop \cite{Borthakur:2007} using the Terasort benchmark
supplied with Hadoop.  Using a benchmark we developed call Terasplit
that computes a single split in a classification and regression tree,
we found that Sector/Sphere was about 1.6--1.9 times faster than
Hadoop.

\section*{Acknowledgments}

This work was supported in part by the National Science Foundation
through grants SCI-0430781, CNS-0420847, and ACI-0325013.

\vfill\eject

\bibliographystyle{plain}

\end{document}